\documentclass[12pt]{article}

\usepackage{lineno,hyperref}
\usepackage{geometry}
\usepackage{amssymb, graphics, setspace}
\usepackage{pgfplots}
\usepackage{textcomp}
\usepackage[caption=false]{subfig}
\usepackage{indentfirst}
\usepackage{amsmath}
\usepackage{commath}
\usepackage{etoolbox}
\makeatletter
\usepackage{graphicx,bm}
\usepackage{url}
\usepackage{float}
\usepackage{textcomp}
\usepackage{bbold}
\usepackage{verbatim}
\usepackage[utf8]{inputenc}
\usepackage{textcomp}

\usepackage{caption}
\usepackage{authblk}

\usepackage{fourier}

\begin{document}

\title{L\'evy $\alpha$-Stable Model for the Non-Exponential Low-$|t|$ Proton--Proton Differential cross section}

\author[1,2]{\small Tam\'as
 Cs\"{o}rg\H{o} \thanks{tcsorgo@cern.ch}}
\author[2]{S\'andor Hegyi \thanks{hegyi.physics@gmail.com}}
\author[1,2,3]{Istv\'an Szanyi \thanks{szanyi.istvan@wigner.hu}}

\affil[1]{\it MATE Institute of Technology,  K\'aroly R\'obert Campus, M\'atrai \'ut 36,
H-3200 Gy\"ongy\"os,  Hungary}
\affil[2]{\it Wigner Research Center for Physics,
 P.O. Box 49,
 H-1525 Budapest, Hungary}
\affil[3]{\it E\"otv\"os University, Department of Atomic Physics,
 P\'azm\'any P. s. 1/A,  H-1117 Budapest, Hungary}

\date{}

\maketitle

\abstract{
\noindent
It is known that the Real Extended Bialas--Bzdak (ReBB) model describes the proton--proton ($pp$) and proton--antiproton ($p\bar p$) differential cross section data in a statistically non-excludible way,\linebreak i.e.,  with a confidence level greater than or equal to 0.1\% in the center of mass energy range \linebreak 546 GeV $\leq\sqrt{s}\leq 8$ TeV and in the squared four-momentum transfer range 0.37 GeV$^2$ $\leq -t\leq1.2$ GeV$^2$. Considering,  instead of Gaussian, a more general L\'evy $\alpha$-stable shape for the parton distributions of the constituent quark and diquark inside the proton and for the relative separation between them, a generalized description of data is obtained, where the ReBB model corresponds to the $\alpha = 2$ special case. Extending the model to $\alpha < 2$, we conjecture that the validity of the model can be extended to  a wider kinematic range, in particular, to lower values of the four-momentum transfer $-t$. We present the formal L\'evy $\alpha$-stable generalization of the Bialas--Bzdak model and show that a simplified version of this model can be successfully fitted, with $\alpha<2$, to the non-exponential, low $-t$ differential cross section data of elastic proton--proton scattering at $\sqrt{s} = 8$ TeV.
}

\section{Introduction}

The Bialas--Bzdak (BB) model considers the proton as a bound state of a quark and a diquark, $p=(q,d)$ for short \cite{Bialas:2006qf}. The diquark in the proton may also be considered to be a weakly bound state of two  constituent quarks, leading to the $p=(q,(q,q))$ variant of the BB model; however,
in Ref.~\cite{Nemes:2012cp}, it was shown that the $p=(q,(q,q))$ variant of the BB model gives two many diffractive minima, whereas, experimentally, only a single minimum is observed in the differential cross section of proton--proton ($pp$) 
collisions. Thus, in recent studies, in Refs.~\cite{Csorgo:2020wmw,Szanyi:2022ezh}, the $p=(q,d)$ version of the model was utilized.

Originally, the BB model considers Gaussian shapes for the parton distributions of constituent quarks and diquarks inside the proton and for the relative separation between them. By these considerations based on R. J.  Glauber's multiple scattering theory~\cite{Glauber:1970jm,Glauber:1984su}, the
inelastic scattering cross section of protons at a fixed $\sqrt{s}$ energy and a fixed $b$ impact parameter value is constructed and denoted as $\tilde\sigma_{in}(s,\vec b)$. 

The elastic scattering amplitude in the impact parameter representation is written in terms of $\tilde\sigma_{in}(s,\vec b)$ as a solution of the unitarity
equation. The imaginary part of the elastic scattering amplitude is the dominant part, whereas the real part can be considered as a smaller correction. Bialas and Bzdak in Ref.~\cite{Bialas:2006qf} neglected the real part of the amplitude and used a fully imaginary amplitude, 
\begin{equation}\label{eq:im_amplitude}
    \tilde t_{el}(s,\vec b) = i\left(1-\sqrt{1-\tilde\sigma_{in}(s,\vec b)}\right),
\end{equation}
for the calculations of the  scattering cross sections. However, in a model where the amplitude does not have a real part, the characteristic minimum--maximum region of the $pp$
differential cross section can not be described properly. In Ref.~\cite{Nemes:2015iia},
the elastic scattering amplitude was extended with a real part in a way that the unitarity constraint is fulfilled. This amplitude reads as:
\begin{equation}\label{eq:uBB_ansatz_sub}
t_{el}(s,\vec b)=i\left(1-e^{i\, \alpha\, \tilde\sigma_{in}(s,\vec b)}\sqrt{1-\tilde\sigma_{in}(s,\vec b)}\right),
\end{equation}
where $\alpha$ is a free parameter to be fitted to the data. In the case of $\alpha = 0$, 
Equation~(\ref{eq:uBB_ansatz_sub}) reduces to Equation~(\ref{eq:im_amplitude}), i.e., to a scattering amplitude that has a vanishing real part.

The model for the elastic proton--proton scattering amplitude, as defined by  \linebreak Equation~(\ref{eq:uBB_ansatz_sub}), with $\tilde\sigma_{in}(s,\vec b)$, as defined in Ref.~\cite{Bialas:2006qf}, is called the Real Extended Bialas--Bzdak (ReBB) model. In recent studies \cite{Csorgo:2020wmw,Szanyi:2022ezh}, it was shown that the ReBB model describes $pp$ and proton--antiproton ($p\bar{p}$) differential cross section data in the center of mass energy range of 0.546~TeV $\leq\sqrt{s}\leq 8$ TeV and in the
squared four-momentum transfer range of  0.37 GeV$^2$ $\leq -t\leq1.2$ GeV$^2$ in a statistically non-excludible manner, i.e., with a confidence level greater than or equal to 0.1\%. 

The free parameters of the ReBB model are the Gaussian radii of the quark, the diquark, and the separation between them (correspondingly, $R_q$, $R_d$, and $R_{qd}$) and also, the $\alpha$ parameter regulating the real part of the scattering amplitude. Two additional fit parameters could be present: $\lambda$, the ratio of the quark and diquark masses, and $A_{qq}$, the normalization parameter appearing in the inelastic quark--quark cross section. However, it was shown in Ref.~\cite{Nemes:2012cp} and later confirmed in Ref.~\cite{Csorgo:2020wmw} that $A_{qq}$ can be fixed at a value of 1.0, whereas $\lambda$ can be fixed at a value of 1/2.

The energy dependence of the ReBB model parameters for $pp$ and $p\bar{p}$ scattering were determined in Ref.~\cite{Csorgo:2020wmw}. It was found that the energy dependencies of the radius parameters are the same for $pp$ and $p\bar{p}$ scattering, whereas the energy dependencies of the $\alpha$ parameter for $pp$ and $p\bar{p}$ scattering are different, i.e., there are different $\alpha^{pp}$ and $\alpha^{p\bar{p}}$ parameters. The energy dependencies of all the five parameters in the energy range of $0.546 \leq \sqrt{s} \leq 8 $ TeV are determined by linear logarithmic functions \cite{Csorgo:2020wmw,Szanyi:2022ezh}.

Considering, instead of Gaussian, a more general L\'evy $\alpha$-stable shape for the parton distributions of the constituent quark and diquark
inside the proton and for the relative separation between them, an improved description to the data in a wider kinematic range ($\sqrt{s}<0.546$~TeV, $\sqrt{s}> 8$ TeV, $-t<0.37$ GeV$^2$, $-t>1.2$ GeV$^2$) is anticipated.

The 0.37 GeV$^2$ $\leq -t\leq1.2$ GeV$^2$ interval at LHC energies includes the region of the characteristic minimum--maximum structure of the $pp$ elastic differential cross section. In the 0.01 GeV$^2$ $\lesssim -t\lesssim0.15$ GeV$^2$  interval, another characteristic structure, a non-exponential behavior is observed. A significant non-exponential behavior was measured by TOTEM at CERN LHC at 8 and 13 TeV center of mass energies \cite{TOTEM:2015oop,TOTEM:2018hki}. Similar behavior was observed also at the CERN ISR accelerator in the 1970s \cite{BARBIELLINI1972663}, where measurements were made in the 20 GeV $\lesssim \sqrt{s}\lesssim60$ GeV energy region.

In Ref.~\cite{Csorgo:2018uyp}, the model-independent L\'evy imaging method is successfully employed to describe the $pp$ and $p\bar p$ differential cross section data both at the low and the high $-t$ region simultaneously. In Ref.~\cite{Csorgo:2019egs}, the model-independent L\'evy imaging method was employed  to reconstruct the proton inelasticity profile function. This method established a statistically significant proton hollowness effect \cite{Dremin:2014dea,Dremin:2014spa,Albacete:2016pmp, Troshin:2017ucy,Broniowski:2018xbg}, well beyond the 5$\sigma$ discovery limit at $\sqrt{s} = 13$~TeV. These results suggest that L\'evy $\alpha$-stable models are efficient tools in describing $pp$ and $p\bar p$ differential cross section data, and the ReBB model needs to be L\'evy $\alpha$-stable generalized to have a stronger non-exponential feature at low $-t$ and to accommodate the new features of the differential cross section data such the hollowness effect at $\sqrt{s} = 13$ TeV or larger energies. In the present work, we complete the formal L\'evy $\alpha$-stable generalization of the Bialas--Bzdak model.

This paper is organized as follows. In Section~\ref{sec:generalization}, we deduce the formal L\'evy $\alpha$-stable generalization of the Bialas--Bzdak model and discuss the technical difficulties preventing us to perform an efficient fitting procedure of the model parameters to the experimental data with the full L\'evy $\alpha$-stable generalized Bialas-Bzdak model. In Section~\ref{sec:simplelevy}, we show successful fits to the low $-t$ differential cross section data at LHC energies with a simple L\'evy $\alpha$-stable model deduced by approximations from the L\'evy $\alpha$-stable generalized Bialas--Bzdak (LBB) model. In Section \ref{sec:connection}, the parameters of the LBB model is related to the $t=0$ measurable quantities and to the parameters of the simple Lévy $\alpha$-stable model.
Finally, we summarize and conclude in Section~\ref{sec:summary}.

\section{From Gaussian to L\'evy \boldmath{$\alpha$}-Stable \boldmath{$p=(q,d)$} BB Model}\label{sec:generalization}

First, we recapitulate the BB model using normalized Gaussian distributions and introduce some reinterpretations of some of its parts. Then, we change the normalized Gaussian distributions to normalized L\'evy $\alpha$-stable distributions, resulting in the L\'evy $\alpha$-stable generalized BB model. 

The inelastic scattering cross section at a fixed $\vec b$ impact parameter value is given as \cite{Bialas:2006qf}:
\begin{equation}\label{eq:tilde_sigma_inel}
\tilde\sigma_{in}(\vec b)=\int_{-\infty}^{+\infty}...\int_{-\infty}^{+\infty} d^{2}\vec s_{q}d^{2}\vec s_{q}^{\,\prime}d^{2}\vec s_{d}d^{2}\vec s_{d}^{\,\prime
}D(\vec s_{q},\vec s_{d})D(\vec s_{q}^{\,\prime},\vec s_{d}^{\,\prime})\sigma(\vec s_{q},\vec s_{d}%
;\vec s_{q}^{\,\prime},\vec s_{d}^{\,\prime};\vec b),
\end{equation}
where $D(\vec s_{q}^{\,\prime},\vec s_{d}^{\,\prime})$ is the quark--diquark distribution inside one of the colliding protons, $\sigma(\vec s_{q},\vec s_{d}%
;\vec s_{q}^{\,\prime},\vec s_{d}^{\,\prime};\vec b)$ is the probability of inelastic collision, and the variables we integrate over are the transverse positions of the quarks and diquarks inside the two colliding protons. Note that the energy dependence of $\tilde\sigma_{in}(\vec b)$ is not written out here for clarity reasons; however, through the $\sqrt{s}$
dependence of the model parameters, $R_q(s)$, $R_d(s)$, and $R_{qd}(s)$, $\tilde\sigma_{in}(\vec b)$ has an $\sqrt{s}$ dependence too.

The quark--diquark distribution is considered to be Gaussian: 
\begin{equation}
D\left({\vec s}_q,{\vec s}_d\right)=\frac{1+\lambda^2}{R_{qd}^2\,\pi}e^{-(s_q^2+s_d^2)/R_{qd}^2}\delta^2({\vec s}_d+\lambda{\vec s}_q),
\label{eq:quark_diquark_distribution}
\end{equation}
where $\lambda=m_q/m_d$, the ratio of the quark and diquark masses, and  $R_{qd}$ are free parameters of the model. The two-dimensional Dirac $\delta$ function fixes the center-of-mass of the proton and reduces the dimension of the integral in Equation~(\ref{eq:tilde_sigma_inel}) from 8 to 4. Accordingly, the diquark positions can be expressed by that of the quarks:
\begin{equation}
{\vec s}_d=-\lambda\,{\vec s}_q,\,\; {\vec s}^{\,\prime}_{d}=-\lambda\,{\vec s}^{\,\prime}_{q}\,.
\end{equation}

{After integration over $\vec s_d$,
$D(\vec s_q, \vec s_d)$ becomes a Gaussian in $\vec s_q$; then, after the integration, also over $\vec s_q$, we obtain unity:
\begin{eqnarray}
 \int d^2\vec s_d D(\vec s_{q},\vec s_{d}) & =& G\left(\vec s_{q} |R_{qd}/\sqrt{2(1+\lambda^2)}\right), \\
 \int d^2\vec s_q d^2\vec s_d D(\vec s_{q},\vec s_{d}) & =& 1,
\end{eqnarray} 
where:
\begin{equation}
    G(\vec x | R_G)= \frac{1}{(2\pi)^2}\int d^2 q e^{i{\vec q}^T \vec x}e^{-\frac{1}{2}q^2R_G^2}=\frac{1}{2\pi R_G^2}e^{-\frac{x^2}{2 R_G^2}}
\end{equation}
is the normalized bivariate Gaussian distribution.}

{We may reinterpret $D(\vec s_q, \vec s_d)$
as the distribution of the relative separation
between the quark and the diquark in a single proton, namely:
\begin{equation}\label{eq:quark-diquark_dist}
    D(\vec s_q, \vec s_d) = (1+\lambda)^2 G\left(\vec s_{q} - \vec s_d | R_{qd}/\sqrt{2}\right)\delta^2(\vec s_d+\lambda \vec s_q)
\end{equation}
which is correctly normalized as follows:
\begin{eqnarray} 
 \int d^2\vec s_d D(\vec s_{q},\vec s_{d}) & =& G\left(\vec s_{q} |R_{qd*}/\sqrt{2}\right), \\
 \int d^2\vec s_q d^2\vec s_d D(\vec s_{q},\vec s_{d}) & =& 1,
\end{eqnarray}
where:
\begin{equation}
R_{qd*}=\frac{R_{qd}}{1+\lambda}.
\end{equation}

Here, we have rescaled the parameter $R_{qd}$ of the original Bialas--Bzdak model to 
the parameter that characterizes the uncertainty of the location of a dressed quark inside the proton. The advantage of this interpretation is that we prepare the ground for the generalization to the case of
Levy $\alpha$-stable distributions and instead of taking the product of two Gaussians, as in Equation~(\ref{eq:quark_diquark_distribution}), we had an equivalent rewrite
where the relative coordinate distribution of a quark and a diquark is Gaussian, with rescaled parameters.
This rewrite is very advantageous, as the product of two Levy distributions is not a Levy distribution,
with the exception of the $\alpha_L = 2$ Gaussian case. As such, to have only one Gaussian in the relative coordinate
avoids the problem of having products of Levy $\alpha$-stable distributions in the formulas.}


The term $\sigma(\vec s_{q},\vec s_{d}%
;\vec s_{q}^{\,\prime},\vec s_{d}^{\,\prime};\vec b)$ is the probability of inelastic
interactions at a fixed impact parameter and transverse positions
of all constituents and given by a Glauber expansion as
follows:
\begin{align}\label{eq:elprob}
\sigma(\vec s_{q},\vec s_{d}%
;\vec s_{q}^{\,\prime},\vec s_{d}^{\,\prime};\vec b)&=1-\left[1-\sigma_{qq}(\vec s_{q},\vec s_{q}^{\,\prime}%
;\vec b)\right]\left[1-\sigma_{qd}(\vec s_{q},\vec s_{d}^{\,\prime}%
;\vec b)\right]\times\\\nonumber &\times\left[1-\sigma_{dq}(\vec s_{q}^{\,\prime},\vec s_{d}%
;\vec b)\right]\left[1-\sigma_{dd}(\vec s_{d},\vec s_{d}^{\,\prime}%
;\vec b)\right],
\end{align}
where:
$$\sigma_{qq}(\vec s_{q},\vec s_{q}^{\,\prime}%
;\vec b)\equiv\sigma_{qq}(\vec b+\vec s_q^{~\prime}-\vec s_q),$$
$$\sigma_{qd}(\vec s_{q},\vec s_{d}^{\,\prime}%
;\vec b)\equiv\sigma_{qd}(\vec b+\vec s_d^{~\prime}-\vec s_q),$$
$$\sigma_{dq}(\vec s_{d},\vec s_{q}^{\,\prime}%
;\vec b)\equiv\sigma_{dq}(\vec b+\vec s_q^{~\prime}-\vec s_d), $$
and:
$$\sigma_{dd}(\vec s_{d},\vec s_{d}^{\,\prime}%
;\vec b)\equiv\sigma_{dd}(\vec b+\vec s_d^{~\prime}-\vec s_d)$$
are the inelastic differential
cross sections of the binary collisions of the constituents. They have Gaussian shapes:
\begin{equation}\label{eq:inelastic_cross_sections}
     \sigma_{ab}(\vec x) = A_{ab}e^{-\vec s^2/S^2_{ab} }
 \end{equation}
with $S^2_{ab}=R_a^2+R_b^2$ and $a,b\in\{q,d\}$. Equation~(\ref{eq:inelastic_cross_sections})  can be rewritten  in terms of normalized bivariate Gaussian distribution:{
\begin{equation}\label{eq:inelastic_cross_sections_NG}
     \sigma_{ab}(\vec s) =  A_{ab} \pi S_{ab}^{2} G\left(\vec s|S_{ab}/\sqrt{2}\right).
 \end{equation}
}

We can reinterpret the inelastic constituent--constituent collisions by assuming that the constituent quark and the constituent diquark have Gaussian parton distributions, characterized by $G(\vec s_q|R_q/\sqrt{2})$ and $G(\vec s_d |R_d/\sqrt{2})$. Then, the probability of inelastic  collisions at a given impact parameter $b$ is proportional to their convolution:{
\begin{align}
         \sigma_{ab}(\vec s) &= A_{ab} \pi S_{ab}^{2}  \int d^2  s_a G(\vec s_a |R_a/\sqrt{2}) G(\vec s - \vec s_a | R_b/\sqrt{2} ) \\\nonumber
         &\equiv  A_{ab} \pi S^{2}_{ab} G\left(\vec s| S_{ab}/\sqrt{2}\right).
\end{align}
}

The inelastic quark--quark, quark--diquark, and diquark--diquark cross sections are obtained by integration:{
\begin{equation}
\label{eq:totalinelastic}
\sigma_{ab,\text{inel}}=\int\limits^{+\infty}_{-\infty}\int\limits^{+\infty}_{-\infty}{\sigma_{ab}\left({\vec s}\right)}\,\text{\rm d}^2s= A_{ab} \pi S_{ab}^{2}\,.
\end{equation}
}

The number of the free parameters of the model can be reduced by demanding that the ratios of the cross sections are:
\begin{equation}
\sigma_{qq,\text{inel}}:\sigma_{qd,\text{inel}}:\sigma_{dd,\text{inel}}=1:2:4\,, 
\label{eq:ratiosforsigma}
\end{equation}
expressing the idea that the constituent diquark contains twice as many partons than the constituent quark and also that the colliding constituents do not ``shadow'' each other.

Then, the probabilities of inelastic constituent--constituent collisions can be written in the following form:{
\begin{equation}
   \sigma_{qq}(\vec s_{q},\vec s_{q}^{\,\prime}%
;\vec b)= 2\pi A_{qq} R_{q}^2 G(\vec b+\vec s_q^{~\prime}-\vec s_q|  R_{q}),
\end{equation}
\begin{equation}
 \sigma_{qd}(\vec s_{q},\vec s_{d}^{\,\prime}%
;\vec b)= 4\pi A_{qq}R_{q}^2 G\Bigg(\vec b+\vec s_d^{~\prime}-\vec s_q\Bigg|\sqrt{\frac{R_{q}^2+R_{d}^2}{2}}\Bigg),
\end{equation}
\begin{equation}
\sigma_{dq}(\vec s_{q}^{\,\prime},\vec s_{d}%
;\vec b) = 4\pi A_{qq}R_{q}^2 G\Bigg(\vec b+\vec s_q^{~\prime}-\vec s_d\Bigg|\sqrt{\frac{R_{q}^2+R_{d}^2}{2}}\Bigg),
\end{equation}
\begin{equation}
\sigma_{dd}(\vec s_{d},\vec s_{d}^{\,\prime}%
;\vec b) = 8\pi A_{qq}R_{q}^2 G(\vec b+\vec s_d^{~\prime}-\vec s_d|R_{d}).
\end{equation}
}

Substituting these into Equation~(\ref{eq:elprob}), then substituting $\sigma(\vec s_{q},\vec s_{d};\vec s_{q}^{\,\prime},\vec s_{d}^{\,\prime};\vec b)$ into Equation~(\ref{eq:tilde_sigma_inel}), we get a sum of eleven integral terms (with proper sign) for $\tilde\sigma_{in}(\vec b)$:
\begin{align}\label{eq:sig_tilde_b}
\tilde\sigma_{in}(\vec b)&=\tilde\sigma^{qq}_{in}(\vec b)+2\tilde\sigma^{qd}_{in}(\vec b)+\tilde\sigma^{dd}_{in}(\vec b)-[2\tilde\sigma^{qq,qd}_{in}(\vec b) +\tilde\sigma^{qd,dq}_{in}(\vec b)+\tilde\sigma^{qq,dd}_{in}(\vec b)+\\ \nonumber
&+2\tilde\sigma^{qd,dd}_{in}(\vec b)]
+[\tilde\sigma^{qq,qd,dq}_{in}(\vec b)+2\tilde\sigma^{qq,qd,dd}_{in}(\vec b)+\tilde\sigma^{dd,qd,dq}_{in}(\vec b)]-\tilde\sigma^{qq,qd,dq,dd}_{in}(\vec b).
\end{align}

Let us have a look for the most general fourth-order term, $\tilde\sigma^{qq,qd,dq,dd}_{in}(\vec b)$. After making use of the presence of the Dirac $\delta$ function in Equation~(\ref{eq:quark-diquark_dist}), we have to calculate a four-dimensional integral of products of normalized bivariate Gaussian distributions:{
\begin{align}\label{eq:tildesigma_general}
\tilde\sigma^{qq,qd,dq,dd}_{in}(\vec b) &= \int d^2s_qd^2s'_qG(\vec s_{q} |R_{qd*}/\sqrt{2})G(\vec s_{q}^{~\prime} |R_{qd*}/\sqrt{2}) \times \\\nonumber &\times \sigma_{qq}(\vec s_{q},\vec s_{q}^{~\prime}%
;\vec b)
\sigma_{qd}(\vec s_{q}, -\lambda \vec s_{q}^{~\prime}%
;\vec b) \sigma_{dq}(\vec s_{q}^{~\prime},-\lambda\vec s_{q}%
;\vec b)\sigma_{dd}(-\lambda \vec s_{q},-\lambda\vec s_{q}^{~\prime}%
;\vec b).
\end{align}
}

Such an integral results in an expression having a Gaussian shape. The lower-order terms can be obtained from Equation~(\ref{eq:tildesigma_general}) by excluding the proper $\sigma_{ab}$ term/terms from the integrand. Thus, after computing the integrals in all order, we get the sum of eleven different Gaussian-shaped terms, i.e., the BB model as introduced in Ref.~\cite{Bialas:2006qf}.

Now, we perform the L\'evy $\alpha$-stable generalization of the BB model.

Let us introduce the normalized bivariate symmetric L\'evy $\alpha$-stable distribution, 

\begin{equation}
L(\vec x|\alpha_L, R_L) = \frac{1}{(2\pi)^2}\int d^2 q e^{i{\vec q}^T \vec x}e^{-\left|q^2R_L^2\right|^{\alpha_L/2}},
\end{equation}
which, for $\alpha_L=2$, gives exactly the bivariate Gaussian distribution: 
\begin{equation}
    L(\vec x|\alpha_L=2, R_L=R_G/\sqrt{2})\equiv G(\vec x | R_G).
\end{equation}
Note that the L\'evy index of stability $\alpha_L$, that controls the power-law tails of the
inelastic cross sections, is a different parameter from the
$\alpha$ parameter of the ReBB model, that controls the opacity or the real part of the 
scattering amplitude. Due to historic reasons, both were denoted by $\alpha$ originally, but in this work, we add a subscripted $L$ to distinguish the L\'evy parameter 
$\alpha_L$ from the opacity parameter $\alpha$.

Since we work with here with symmetric L\'evy $\alpha$-stable distribution, the skewness parameter $\beta_L = 0$ of the L\'evy stable source distributions is  implicit and are assumed to have zero values. The shift parameter $\delta_L$ of the L\'evy stable source distribution is explicitely written out when considering the impact parameter picture, while the overall shift
of the impact parameter cancels from the final results hence it is assumed to have a vanishing value.

We then consider that the relative separation
between the quark and the diquark in a single proton follows L\'evy $\alpha$-stable distribution:
\begin{equation}\label{eq:quark-diquark_levy}
    D(\vec s_q, \vec s_d) = (1+\lambda)^2 L\left(\vec s_{q} - \vec s_d | \alpha_L, R_L =  R_{qd}/2\right) \delta^2(\vec s_d+\lambda \vec s_q)
\end{equation}
with:
\begin{eqnarray} 
 \int d^2\vec s_d D(\vec s_{q},\vec s_{d}) & =& L\left(\vec s_{q} |\alpha_L,R_{qd*}/2\right), \\
 \int d^2\vec s_q d^2\vec s_d D(\vec s_{q},\vec s_{d}) & =& 1,
\end{eqnarray}

similarly to the original case with Gaussian distributions.

As the next step in the generalization, we consider, instead of Gaussian,  L\'evy $\alpha$-stable parton distributions for the constituent quark and the constituent diquark: $L(\vec s_q|\alpha_L, R_q/2)$ and $L(\vec s_d |\alpha_L, R_d/2)$. 
Then, as in the Gaussian case above, the probability of inelastic  collisions at a given impact parameter $b$ is proportional to their convolution: {
\begin{align}
         \sigma_{ab}(\vec s) &= A_{ab} \pi S_{ab}^{2}   \int d^2  s_a L(\vec s_a |\alpha_L,R_a/2) L(\vec s - \vec s_a | \alpha_L, R_b/2 ) \\\nonumber
         &=  A_{ab} \pi S_{ab}^{2} L\left(\vec s|\alpha_L, S_{ab}/2\right),
\end{align}}
where now:
\begin{equation}
    S_{ab}= \left(R_a^{\alpha_L}+R_b^{\alpha_L}\right)^{1/\alpha_L},
\end{equation}

i.e., in this case, after making use of the convolution theorem, the radii add up not quadratically, but at the power of $\alpha_L$.

Then{:
\begin{equation}\label{eq:qqlevy}
\sigma_{qq}(\vec s_{q},\vec s_{q}^{\,\prime}%
;\vec b)= \pi A_{qq} \left(2R_{q}^{\alpha_L}\right)^{2/\alpha_L} L\left(\vec b+\vec s_q^{~\prime}-\vec s_q|\alpha_L,  \left(2 R_{q}^{\alpha_L}\right)^{1/\alpha_L}/2\right),
\end{equation}
\begin{equation}\label{eq:qdlevy}
\sigma_{qd}(\vec s_{q},\vec s_{d}^{\,\prime}%
;\vec b)= 2\pi A_{qq}\left(2R_{q}^{\alpha_L}\right)^{2/\alpha_L} L\Bigg(\vec b+\vec s_d^{~\prime}-\vec s_q\Bigg|\alpha_L,\left(R_q^{\alpha_L}+R_d^{\alpha_L}\right)^{1/\alpha_L}/2\Bigg),
\end{equation}
\begin{equation}\label{eq:dqlevy}
\sigma_{dq}(\vec s_{q}^{\,\prime},\vec s_{d}%
;\vec b) = 2\pi A_{qq}\left(2R_{q}^{\alpha_L}\right)^{2/\alpha_L} L\Bigg(\vec b+\vec s_q^{~\prime}-\vec s_d\Bigg|\alpha_L,\left(R_q^{\alpha_L}+R_d^{\alpha_L}\right)^{1/\alpha_L}/2\Bigg),
\end{equation}
and
\begin{equation}\label{eq:ddlevy}
\sigma_{dd}(\vec s_{d},\vec s_{d}^{\,\prime}%
;\vec b) = 4\pi A_{qq}\left(2R_{q}^{\alpha_L}\right)^{2/\alpha_L} L\left(\vec b+\vec s_d^{~\prime}-\vec s_d)|\alpha_L,\left(2 R_{d}^{\alpha_L}\right)^{1/\alpha_L}/2\right).
\end{equation}
}

Equation~(\ref{eq:tilde_sigma_inel}) with Equation~(\ref{eq:elprob}), Equation~(\ref{eq:quark-diquark_levy}), and Equations~(\ref{eq:qqlevy})--(\ref{eq:ddlevy}) define the L\'evy $\alpha_L$-stable generalized Bialas--Bzdak (LBB) model for $\tilde\sigma_{in}(b)$. Now, in Equation~(\ref{eq:sig_tilde_b}), instead of a sum of integrals of products of normalized Gaussian distributions, there are a sum of integrals of products of normalized L\'evy $\alpha_L$-stable distributions. Though integrals of products of Gaussian distributions can be calculated, the calculation of integrals of products of L\'evy $\alpha_L$-stable distributions is an issue. Integrals of products of L\'evy $\alpha_L$-stable distributions can be easily calculated if the integral can be written in a convolution form. This is the case for the first three terms in Equation~(\ref{eq:sig_tilde_b}). The results can be written in terms of L\'evy $\alpha_L$-stable distributions: 

{
\begin{align}\label{eq:LBBsqqb}
\tilde\sigma_{in}^{qq}(\vec b)&=  \pi A_{qq} \left(2R_{q}^{\alpha_L}\right)^{2/\alpha_L}\times\\ \nonumber &\times\int d^2s_qd^2s'_qL(\vec s_{q} |\alpha_L, R_{qd*}/2)L(\vec s_q^{~\prime} |R_{qd*}/2)L\left(\vec b+\vec s_q^{~\prime}-\vec s_q|\left(2 R_{q}^{\alpha_L}\right)^{1/\alpha_L}/2\right) \\ \nonumber 
&=\pi A_{qq} \left(2R_{q}^{\alpha_L}\right)^{2/\alpha_L} L\left(\vec b\Big|\alpha_L,\left(2R_{qd*}^{\alpha_L}+2R_{q}^{\alpha_L}\right)^{1/\alpha_L}/2\right),
\end{align}
\begin{align}\label{eq:LBBsqdb}
\tilde\sigma_{in}^{qd}(\vec b)&=2\pi A_{qq} \left(2R_{q}^{\alpha_L}\right)^{2/\alpha_L}\times\\ \nonumber &\times\int d^2s_qd^2s'_qL(\vec s_{q} |R_{qd*}/2)L(\vec s_{q}^{~\prime} |R_{qd*}/2)L\left(\vec b-\lambda\vec s_q^{~\prime}-\vec s_q\Bigg|\alpha_L,\left(R_{q}^{\alpha_L}+R_{d}^{\alpha_L}\right)^{1/\alpha_L}/2\right)\\ \nonumber 
&= 
2\pi A_{qq} \left(2R_{q}^{\alpha_L}\right)^{2/\alpha_L} L\left(\vec b\Big|\alpha_L,\left((1+\lambda^{\alpha_L})R_{qd*}^{\alpha_L}+ R_{q}^{\alpha_L}+R_{d}^{\alpha_L}\right)^{1/\alpha_L}/2\right),
\end{align}
\begin{align}\label{eq:LBBsddb}
\tilde\sigma_{in}^{dd}(\vec b)
&=4\pi A_{qq} \left(2R_{q}^{\alpha_L}\right)^{2/\alpha_L}\times\\ \nonumber &\times\int d^2s_qd^2s'_qL(\vec s_{q} |R_{qd*}/2)L(\vec s_q^{~\prime} |R_{qd*}/2)L\left(\vec b+\lambda(\vec s_q-\vec s_q^{~\prime})|\alpha_L,\left(2 R_{d}^{\alpha_L}\right)^{1/\alpha_L}/2\right) \\ \nonumber
&=4\pi A_{qq} \left(2R_{q}^{\alpha_L}\right)^{2/\alpha_L} L\left(\vec b\Big|\alpha_L,\left(2\lambda^{\alpha_L} R_{qd*}^{\alpha_L}+2R_{d}^{\alpha_L}\right)^{1/\alpha_L}/2\right).
\end{align}
}

The results of the remaining eight integrals, corresponding to higher-order terms in the BB model, are yet to be determined in terms of analytic formulas.

Whereas univariate and multivariate Gaussian distributions have closed forms in terms of elementary functions, univariate and multivariate L\'evy $\alpha_L$-stable distributions have forms in terms of special functions. 
This makes it hard to perform a numerical fitting procedure of the model parameters to the experimental data. To complete this work in the future, a relatively high computing capacity or improved analytic insight  will be needed. In this work, we have chosen another approach, limiting the domain of the applicability of the calculations in the squared four-momentum transfer $-t$. This allows for certain simplifications and results in an increased analytic insight to certain properties of the LBB~model.

{A possible alternative to the L\'evy $\alpha$-stable generalization of the BB model could be its Tsallis or q-exponential generalization, since data from high-energy collisions have shown such distribution. The presence of the Tsallis distribution was explained in Ref.~\cite{Deppman:2019yno} using the fractal approach to the non-perturbative QCD, and also, the $q$ index was expressed in terms of the number of colors and the number of flavors. The validity of the derived relation was reinforced later in Ref.~\cite{Megias:2023hyh}. These results suggest that the investigation of the Tsallis generalization of the BB model is worthwhile. This will be done in a future study. In this manuscript, we investigate the L\'evy $\alpha$-stable generalization of the BB model.}

\section{A Simple L\'evy \boldmath{$\alpha$}-Stable Model}\label{sec:simplelevy}

Now, we check if the L\'evy $\alpha$-stable generalization of the BB model has an enhanced potential, as compared to the ReBB model, or not. The mathematical and computing difficulties discussed in the previous section can be bypassed by introducing new approximations that are valid at low $-t$, in the domain where the original ReBB model had difficulties to describe the strongly non-exponential features of the experimental data of elastic proton-proton scattering at the TeV energy scale. Our aim is, thus, to deduce a model for the differential cross section which is valid at the low-$|t|$ region.

{Low-$|t|$ scattering corresponds to high-$b$ scattering and, at high $b$ values $\tilde\sigma_{in}(s,b)$, is small. 
Thus, the leading order term in the Taylor expansion of Equation~(\ref{eq:uBB_ansatz_sub}), i.e.,
\begin{equation}\label{eq:full_amplitude_expanded}
    \tilde t_{el}(s,\vec b) = \left(\alpha+\frac{i}{2}\right)\tilde\sigma_{in}(s,\vec b),
\end{equation}
should be a reasonable approximation at low $-t$ values if the $\alpha$ parameter controlling the real part of the scattering amplitude is small. Since in this approximation 
\begin{equation}\label{eq:rho0_ReBB}
    \rho_{0}(s)=\frac{Ret(s,t=0)}{Imt(s,t=0)}=2\alpha(s),
\end{equation}
one can rewrite\footnote{Note added for arXiv.org version.} Eq.~(\ref{eq:full_amplitude_expanded}) as
\begin{equation}\label{eq:full_amplitude_expanded_rewrite}
    \tilde t_{el}(s,\vec b) = \frac{1}{2}\left(i+\rho_0(s)\right)\tilde\sigma_{in}(s,\vec b).
\end{equation}

In Section~\ref{sec:generalization}, we discussed that in the L\'evy $\alpha$-stable generalized case of the BB model, the leading order terms in $\tilde\sigma_{in}(s,b)$ are L\'evy-$\alpha$-stable-shaped terms. Motivated by this fact in our simplified model, we approximate $\tilde\sigma_{in}(s,b)$ with a single  L\'evy-$\alpha$-stable-shaped term, i.e.,
\begin{equation}\label{eq:sigmain_levy}
\tilde\sigma_{in}(s,\vec b)=\tilde c(s) L(\vec b |\alpha_L(s),  r(s))
\end{equation}
where $\tilde c(s)$,  
$\alpha_L(s)$, and $ r(s)$ are energy dependent parameters.

Then, by Equation~(\ref{eq:full_amplitude_expanded_rewrite}), we have:
\begin{equation}
    \tilde t_{el}(s,\vec b) = c(s) L(\vec b |\alpha_L(s),  r(s)),
\end{equation}
where $ c(s)=\frac{1}{2}\left(i+\rho_0(s)\right)\tilde c(s)$ is a rescaled and complex valued parameter.
Now, we transform the impact parameter amplitude into momentum space:
\begin{equation}\label{eq:im_amplitude_expanded_transformed}
t(s,t)=\int d^2 b e^{i\vec \Delta^T\vec b}\tilde t(s,\vec b) = c(s)e^{-|t r^2(s)|^{\alpha_L(s)/2}},
\end{equation}
where $|{\vec\Delta}|\simeq \sqrt{-t}$. 
The resulting differential cross section is:  
 \begin{equation}
\frac{{\rm d}\sigma}{{\rm d} t}(s,t) = \frac{1}{4\pi}\left|t\left(s,t\right)\right|^2 = a(s)e^{-|t b(s)|^{\alpha_L(s)/2}},
\label{eq:differential_cross_section_final}
\end{equation}
where $a(s)=\frac{|c(s)|^2}{4\pi}=\frac{1}{16\pi}(1+\rho_0^2){\tilde c}^2 (s)$ and $b(s)=2^{2/\alpha_L(s)} r^2(s)$. Since the total scattering cross section is 
\begin{equation}
   \sigma_{tot} (s)=2Im t(s,t=0), 
\end{equation}
it is easy to see that\footnote{Note added for arXiv.org version.} ${\tilde c} (s) = \sigma_{tot} (s)$. 
Aside from all that, we simply use Equation~(\ref{eq:differential_cross_section_final})
as a model for the differential cross sections with three adjustable parameters, $\alpha_L$, $a$, and $b$, to be determined at a given energy.}

\begin{figure}[hbt!]
\centering
\includegraphics[width=0.6\textwidth]{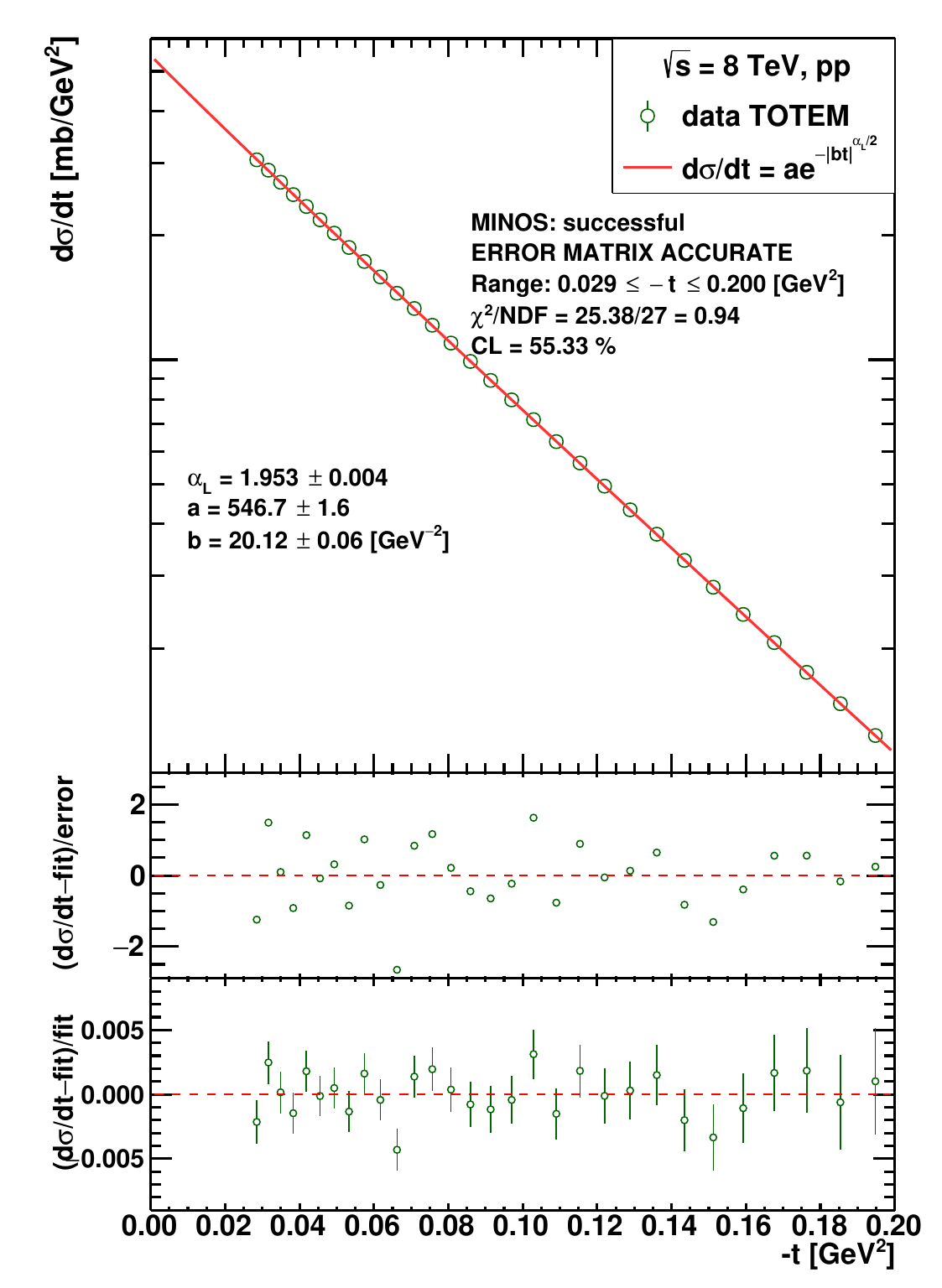}
\caption{ 
Fit-to-the-TOTEM
 $pp$ elastic differential cross section data at  $\sqrt{s}=8$ TeV \cite{TOTEM:2015oop} by the model defined by Equation~(\ref{eq:differential_cross_section_final}). 
}
\label{fig:fig1_8TeV}
\end{figure}

The result of a fit to the TOTEM $pp$ elastic differential cross section data at $\sqrt{s}=8$~TeV by the model defined by Equation~(\ref{eq:differential_cross_section_final}) is shown in Figure~\ref{fig:fig1_8TeV}. One can see that the non-exponential model with $\alpha_L$ = 1.953 $\pm$ 0.004 represents the low-$|t|$ differential cross section data with a confidence level of 55\%.

\begin{figure}[hbt!]
\centering
\includegraphics[width=0.6\textwidth]{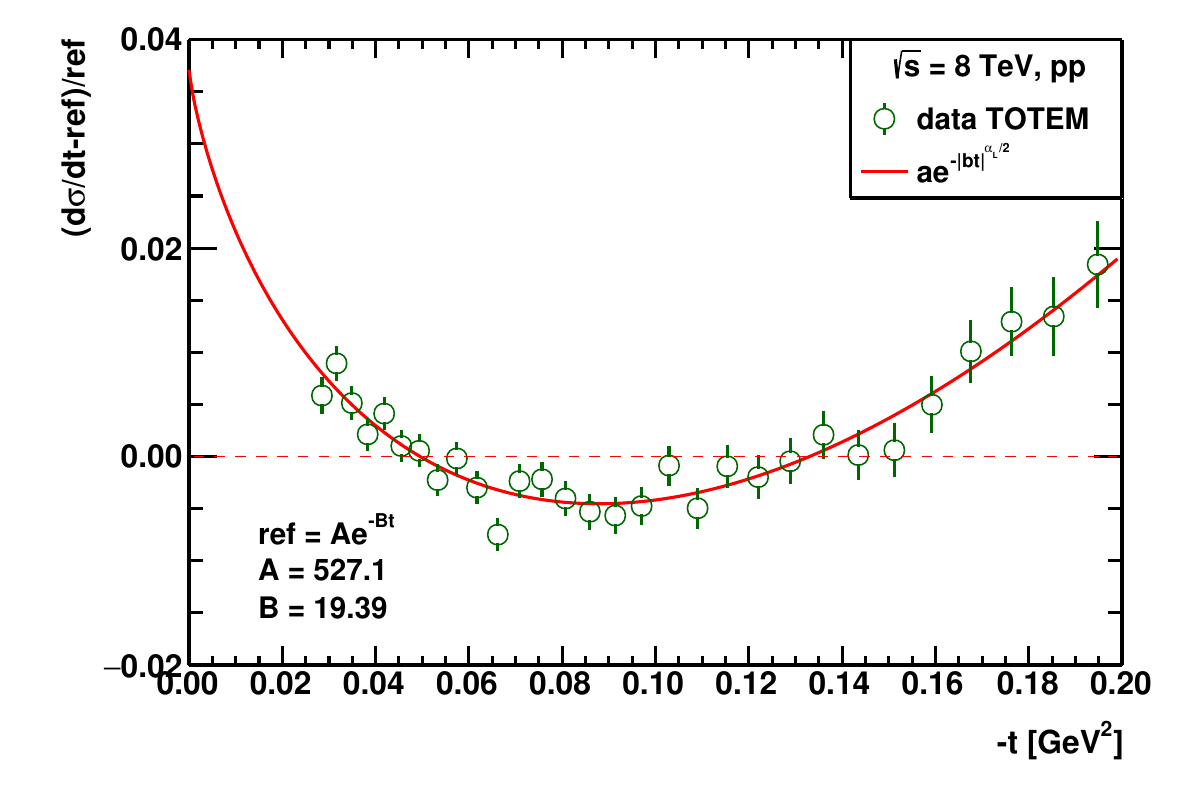}
\caption{The ratio,
 $(d\sigma/dt-ref)/ref$, evaluated from the 
TOTEM $pp$ elastic differential cross section data at  $\sqrt{s}=8$ TeV \cite{TOTEM:2015oop}. The curve corresponds to the fitted model defined by Equation~(\ref{eq:differential_cross_section_final}).
}
\label{fig:fig2}
\end{figure}

Figure~\ref{fig:fig2} shows the ratio, $(d\sigma/dt-ref)/ref$, with $ref=Ae^{-Bt}$,
used by the TOTEM collaboration \cite{TOTEM:2015oop} to make the relatively small, but significant, low-$|t|$ non-exponential behavior visible. One can clearly see that our model successfully describes the low-$|t|$ data.

{
\section{The \boldmath{$t=0$} Measurable Quantities and the BB Model Parameters}\label{sec:connection}

In this section, we relate the $t=0$ measurable quantities and the LBB parameters. First, we work with the original BB model with Gaussian distributions and then derive the formulas for the Levy $\alpha$-stable generalized case.  We note again that to avoid confusion with the $\alpha$ parameter of the ReBB model regulating the real part of the amplitude and that of the Lévy $\alpha$-stable distribution, the latter we have denoted in this manuscript
as $\alpha_L$. 
For the $\alpha_L=2$ limiting case, the relations obtained in the original BB model are recovered.  

For the scattering amplitude, we use the approximation as defined by Equation~(\ref{eq:full_amplitude_expanded}), and we consider only the leading order terms in $\tilde\sigma_{in}(s,\vec b)$, i.e., $\tilde\sigma^{qq}_{in}(\vec b)$, $\tilde\sigma^{qd}_{in}(\vec b)$, and $\tilde\sigma^{dd}_{in}(\vec b)$, which give the dominant contribution at $t=0$. We get the amplitude in momentum space by Fourier transformation as in Equation~(\ref{eq:im_amplitude_expanded_transformed}). As discussed in the Introduction, the parameter $A_{qq}$ can be fixed at a value of 1.0, whereas $\lambda$ can be fixed at a value of 1/2. We use these specific values below. 

With Gaussian distributions in the BB model, in the low-$|t|$ approximation, $\sigma_{tot}$ is related to the square of the quark radius $R_q$,
\begin{equation}\label{eq:sigtot_ReBB}
    \sigma_{tot}=18\pi R_q^2,
\end{equation}
whereas the ratio of the real to the imaginary part of the forward scattering amplitude is related to the $\alpha$ parameter of the ReBB model,$\rho_{0}=2\alpha$, as discussed in Sec.~\ref{sec:simplelevy}. Note that this result for $\rho_{0}$ holds also in the Levy $\alpha$-stable generalized case.

The low-$|t|$ $pp$ differential cross section can be written in the form \cite{TOTEM:2015oop}:
\begin{equation}
   \frac{d\sigma}{dt}= \frac{1}{4\pi}\left|t\left(s,t\right)\right|^2=ae^{-b_1t+b_2t^2}
\end{equation}
where:
\begin{equation}
    a=\frac{d\sigma}{dt}\Big|_{t=0}
\end{equation}
is the optical point,
\begin{equation}
   b_1=\left(\frac{d}{dt}\ln\frac{d\sigma}{dt}\right)\Bigg|_{t=0}
\end{equation}
is the slope parameter,
and 
\begin{equation}
   b_2=\frac{1}{2}\left(\frac{d}{dt^2}\ln\frac{d\sigma}{dt}\right)\Bigg|_{t=0}
\end{equation}
is the curvature parameter.
These measurable quantities can be expressed in terms of the ReBB model parameters:
\begin{equation}\label{eq:op_ReBB}
    a=\frac{81}{4} \pi Rq^4 \left(1 + 4 \alpha^2\right),
\end{equation}
\begin{equation}\label{eq:b1_ReBB}
   b_1=\frac{2}{9}R_{qd}^2+\frac{2}{3}R_{d}^2+\frac{1}{3}R_{q}^2,
\end{equation}
and
\begin{equation}\label{eq:b2_ReBB}
   b_2=\frac{1}{324}\left(R_{qd}^2-3R_{d}^2+3R_{q}^2\right)^2.
\end{equation}

Now, we turn to the LBB model. Using the Levy $\alpha$-stable generalized forms of the leading order terms in $\tilde\sigma_{in}(s,b)$, i.e., Equations~(\ref{eq:LBBsqqb})--(\ref{eq:LBBsddb}), the total cross section is:
\begin{equation}\label{eq:sigtot_LBB}
    \sigma_{tot}=9\pi \left(2R_q^{\alpha_L}\right)^{2/\alpha_L}.
\end{equation}

Furthermore, we consider now that the differential cross section has the form as written in Equation~(\ref{eq:differential_cross_section_final}). Now, the optical point is:
\begin{equation} \label{eq:op_LBB}
   a=\frac{81}{16} \pi \left(2 Rq^{\alpha_L}\right)^{4/\alpha_L} \left(1 + 4 \alpha^2\right),
\end{equation}
whereas the slope parameter is:
\begin{equation}\label{eq:b1_LBB}
   b=\frac{1}{36}\left(\frac{4}{3}\right)^{2/\alpha_L}\left(\left(2+2^{\alpha_L}\right)R_{qd}^{\alpha_L}+3^{\alpha_L}\left(2R_d^{\alpha_L}+R_q^{\alpha_L}\right)\right)^{2/\alpha_L}.
\end{equation}

One can easily check that for $\alpha_L=2$, Equation~(\ref{eq:sigtot_LBB}) reduces to Equation~(\ref{eq:sigtot_ReBB}), \linebreak Equation~(\ref{eq:op_LBB}) to Equation~(\ref{eq:op_ReBB}), and Equation~(\ref{eq:b1_LBB}) to Equation~(\ref{eq:b1_ReBB}). Since the function in Equation~(\ref{eq:differential_cross_section_final}) is not an analytic function of $t$ at $t=0$, Equation~(\ref{eq:b1_LBB}) was obtained by a Taylor expansion in $t^{\alpha_L/2}$ around zero and by keeping only the leading order term.

As discussed in Sec.~\ref{sec:simplelevy}, the Levy scale parameter $r$ in our simple L\'evy $\alpha$-stable model is related to the slope parameter. The relation can be rewritten as $r=\sqrt{b}/2^{1/\alpha_L}$. Then, this $r$ parameter can be expressed in terms of the LBB model parameters:
\begin{equation}
   r=\frac{1}{6}\left(\frac{2}{3}\right)^{1/\alpha_L}\left(\left(2+2^{\alpha_L}\right)R_{qd}^{\alpha_L}+3^{\alpha_L}\left(2R_d^{\alpha_L}+R_q^{\alpha_L}\right)\right)^{1/\alpha_L}.
\end{equation}

Thus, we have shown that the parameters of our simple L\'evy $\alpha$-stable model, namely, $a$ and $b$ (or equivalently, $r$), can be approximately expressed in terms of those of the LBB~model. 

In Ref.~\cite{Petrov:2018ays}, the three-dimensional radius of the proton is defined and its relation to the slope parameter is derived. In our work, we related the Levy scale parameter $r$ in our simple L\'evy $\alpha$-stable model to the elastic slope parameter and expressed it in terms of the radii of the constituents of the proton ($R_q$ and $R_d$) and their typical separation ($R_{qd}$).

Finally, we note that there are five measurable parameters at the forward region: the total cross section, the ratio of the real to the imaginary part of the forward scattering amplitude, the optical point, the slope parameter, and the curvature parameter. The ReBB model has four free parameters, whereas the LBB model has five. This naturally suggests that the LBB can give a better description to the data than the ReBB model.
}

 \section{Summary}\label{sec:summary}
 
The ReBB model turned out to be an efficient tool in  describing $pp$ and $p\bar p$ differential cross section data, but in a limited $\sqrt{s}$ and $-t$ range. The validity range of the ReBB model in $\sqrt{s}$ does not include 13 TeV, possibly due to the significant hollowness effect observed at that energy. The validity range of the ReBB model in $-t$ includes the minimum--maximum structure of the differential cross section, but does not include the significant non-exponential behavior at low $-t$ values. To overcome these shortcomings of the ReBB model, in this paper, we introduce the L\'evy $\alpha$-stable generalized Real Extended Bialas--Bzdak (LBB) model. The fitting of the parameters of the LBB model to the experimental data, however, requires the solution of difficult and complex technical (mathematical and computational) problems. However, in the low four-momentum transfer region,
based on our novel approximations and the idea of the Levy-$\alpha$-stable-shaped inelastic scattering probability suggested by the LBB model, we deduced and fitted a highly simplified Levy $\alpha$-stable model of the $pp$ differential cross section to the measured data at $\sqrt{s}=8$ TeV. The results show that our simple model represents the low-$|t|$ experimental data in a statistically acceptable manner. This is a promising prospect for the future utility of the L\'evy $\alpha$-stable generalized Real Extended Bialas--Bzdak (LBB) model. 

{ We have shown also that the parameters of our simple L\'evy $\alpha$-stable model, namely, $a$ and $b$ (or equivalently, $r$), can be approximately expressed in terms of those of the LBB model, which is based on R. J. Glauber's multiple diffractive scattering theory. We emphasize that there are five measurable parameters at the forward region, whereas the ReBB model has only four free parameters. Since the LBB model has five free parameters, it is natural to expect that it can give a better description to the data than the ReBB model.}

In the next steps of our research, we are planning to extend the fits with our simple model for all the energies where low-$|t|$ experimental data exist, and after solving the technical issues, to fit the full LBB model to all the existing experimental $pp$ and $p\bar p$ differential cross section data.

\section*{Acknowledgents}
The work was supported by the NKFIH Grants no. K133046 and 2020-2.2.1-ED-2021-00181, as well as by the \'UNKP-22-3 New National Excellence Program of the Ministry for Innovation and
	Technology from the source of the National Research, Development and Innovation Fund.

\end{document}